 \documentstyle[aps,multicol]{revtex}
 \input{epsf}
 \begin{document} 
\draft 
\columnsep -.375in 
\title{
Quantum interference and electron-electron interactions\\ 
at strong spin-orbit coupling in disordered systems}
\author{Yuli Lyanda-Geller} 
\address{
Department of Physics, 
Materials Research Laboratory and  
Beckman Institute, 
University of Illinois,
Urbana, IL 61801
}  
\date{11 January 1998}
\maketitle
\begin{abstract} 
Transport and thermodynamic properties of disordered conductors are considerably 
modified when the angle through which the electron spin precesses due to spin-orbit 
interaction (SOI) during the mean free time becomes significant. 
Cooperon and Diffusion equations are solved for the entire range of strength 
of SOI. The implications of SOI for the electron-electron interaction 
and interference effects in various experimental settings are discussed. 
\end{abstract}
\begin{multicols}{2}
\narrowtext
\noindent
{\it Introduction\/}: 
The effects of weak localization (WL) and electron-electron interaction on transport 
in disordered conductors are strongly influenced by interactions that affect 
electron phase coherence:  by magnetic fields, magnetic impurities and spin-orbit 
interactions (SOI).  The issue of the effects of SOI on WL \cite{Abrahams,Gorkov}
and electron-electron interaction effects \cite{Alt1}
attracted considerable attention in early studies of WL corrections 
to conductivity \cite{Larkin,Altshuler,Fukuyama,Lee_Alt,Bergmann}.  More recently 
\cite{ST_Math,LGM,Aronov,ILGP,LG}, it was shown that, in addition, SOI can be 
regarded as generating an effective spin-dependent vector-potential, which 
influences electron coherence rather like the electromagnetic vector potential 
does (via the Aharonov-Bohm, or AB, effect).
To date, quantum corrections to conductivity 
have been conventionally studied under the assumption that the characteristic time scale 
which determines the SOI strength, $\tau_{so}$, significanly exceeds the mean 
free time $\tau$ \cite{Larkin,Altshuler,Fukuyama,Lee_Alt,Bergmann,ST_Math,LGM,Aronov,ILGP,LG,Wheeler,Knap}.

In the present Letter, I discuss quantum transport phenomena associated with SOI of arbitrary
strength.  Systems with strong SOI, $\tau_{so}\le\tau$, are now intensively studied experimentally  
\cite{Kravchenko,Simonian}. Of particular concern here will be
implications of strong SOI for WL and electron-electron
interaction corrections to the conductivity.

It is important to recognize that two types of SOI can be identified.  
First, there is {\it random\/} SOI, due to impurity potentials.
The scattering amplitude contains a spin-independent term and a much smaller spin-dependent term 
which, however, leads to SOI dephasing~\cite{Larkin}. 
The SOI dephasing time due to this random SOI is always much larger than $\tau$.
  The second type of SOI occurs in low-dimensional and low-symmetry systems, and owes 
its existence to the crystalline or confining potential. In this case, the 
electron Hamiltonian has the form
\begin{equation}
{\cal H}=p^2/2m^{*}+
\hbar\mbox{\boldmath $\sigma$}\cdot\mbox{\boldmath$\Omega$}(\mbox{\boldmath $p$}), 
\label{EQ:soi_term}
\end{equation} 
where $m^{*}$ is the effective electron mass, and $\Omega$ can be 
regarded as momentum-dependent 
spin-precession frequency. This type of SOI characterizes several recent 
experimental settings 
\cite{Wheeler,Knap,Kravchenko,Simonian}.  
I consider here forms of ${\bf\Omega}({\bf p})$ that transform like the Legendre polynomial 
$P_{1}$, which characterize two-dimensional (2D) systems (e.g., Si MOSFETs) 
and 1D GaAs quantum wires and rings. Therefore
$\Omega_{i}({\bf p})= \beta_{ij}p_j\,$
and the spin term in Eq.~(\ref{EQ:soi_term}) can be written as
${\bf \sigma}\cdot
{\bf\Omega}({\bf p})=
{\bf p}\cdot{\bf \tilde{A}}/m^{*}, 
$ where ${\bf\tilde{A}}$ is the spin-dependent vector potential~\cite{A}.  
It results in conductance oscillations in quantum rings \cite{ST_Math,Aronov}, 
an unusual random matrix ensemble \cite{LGM}, and anomalous magnetoresistance 
(MR) in 2D structures \cite{ILGP,LG,Wheeler,Knap}. These phenomena can be 
regarded \cite{ST_Math,LG} as 
manifestations of  the Aharonov-Casher (AC) effect~\cite{Aharonov} in disordered 
electronic systems.

The strength of SOI in Eq.~(\ref{EQ:soi_term}) can be characterized, in semiclassical 
terms, by the angle of spin precession during $\tau$, $\Omega\tau$. 
When $\Omega\tau\ll 1$, the 
SOI dephasing time due to ${\tilde{\bf A}}$,
is $1/\langle\Omega^2({\bf p})\rangle\tau\gg\tau, 
$ as for random SOI. For arbitrary $\Omega\tau$~\cite{strength} 
this is no longer the case. 

The main results of this Letter are as follows: 
(i)~At strong SOI positive magnetoresistance persists
in 2D weakly disordered conductors in the whole range 
of magnetic fields. 
(ii)~Due to 
electron-electron interactions, AC oscillations arise in the conductivity, 
the density of states, and thermodynamic quantities. 

\noindent
{\it SOI and the interference correction to conductivity\/}:  
We now address the issue of how SOI of arbitrary strength influences the WL 
correction.  We note, in passing, that 
the classical (i.e., Drude) expression for the conductivity $\sigma_0$ is left 
unchanged by SOI in Eq.~(\ref{EQ:soi_term}), and $
\sigma_{0}=e^2 n\tau /m^*$.
Interference corrections for disordered conductors in the diffusive regime have 
their origin in the increased amplitude for phase-coherent electron propagation 
along self-crossing trajectories.  
In order to address interference corrections to $\sigma_0$, one retains the  
maximally crossed diagramms in the quantity $G_{\epsilon + \omega /2}^{R}({\bf p} + {\bf q}/2)
G_{\epsilon -\omega /2}^{A}({\bf p} - {\bf q}/2)$, where  $G^{R}$ ($G^{A}$) are the single 
electron retarded (advanced) Green functions, ${\bf q}$ is the total momentum of 
particles whose correlation is described, and thereby 
arrives at an equation for the Cooperon propagator (see, for instance, Ref.~\cite{Alt_Ar}).  
Similarly, ladder diagrams give rise to 
the Diffuson equation.  For the physical system under consideration, 
the spin-dependence in the Cooperon/Diffuson equations arises from propagation, 
i.e. results from $G^{R}$ ($G^{A}$), and not from scattering. 
The Cooperon equation is given by: 
\begin{equation}
{\cal C}= 1 + \int \frac{d{\bf o}}{1 + i\omega\tau + 
i{\bf p}\tau({\bf q} + 2e{\bf A}_{\rm em}/c + {\bf A})/m}{\cal C},
\label{EQ:Louiville}
\end{equation}
where ${\bf o}$ denote the orientation of momenta ${\bf p}$, ${\bf A}_{\rm em}$
is the external electromagnetic vector-potential, $\omega$ is the frequency, 
${\bf A}$ is the spin-dependent 
vector potential, $A_j = 2\beta_{ij}{\bf S}_i$,and $S$ is the total spin of particles.
 The conventional approach to Eq.~\ref{EQ:Louiville} 
is the expansion of the integrand up to the second order in $ql$ and ${\bf A}$, 
leading to diffusion-like 
equation for the Cooperon/ Diffuson propagators. In the present Letter, 
we calculate these propagators exactly, 
without such an expansion. 
Consider first a quasi 1D wire lying along the $z$-direction or a quasi 1D ring 
with angular coordinate $\phi$. Let SOI be described by tensor
$\beta$ having nonzero components $\beta_{xx}=\beta_1$ (or $\beta_{\phi\phi}=\beta_1$) 
which is a good approximation for narrow constrictions \cite{narrow}. In this case
the solution of the Eq.~(\ref{EQ:Louiville}) is
\begin{equation} 
{\cal C}_{Sj} = \frac{1}{D\tau (q + 2j\beta m)^2 + i\omega\tau},
\label{EQ:Cooperon/one}
\end{equation}
where $j$ is {\bf S}-projection along ${\bf q}$, $D$ is the diffusion coefficient.
Remarkably, the Eq~(\ref{EQ:Cooperon/one}), derived without 
the ${\bf A}$- and $ql$-expansion, has the same 
form as the Cooperon propagator for weak SOI
in~\cite{LGM}. At the same time, the physical 
properties of systems, described by Eq.~(\ref{EQ:Cooperon/one}), are  
determined by contributions from $q\sim j\beta m^{*}$. If 
SOI is of intermediate strength, i.e. $\beta m^{*}l \sim 1$, such $q$ mean $ql\sim 1$. 
Therefore, strong SOI 
(which is, of course, treated {\it without using expansion} in powers of ${\bf A}$)
cannot be studied properly, in general, if the conventional $ql$-expansion is applied.
 This is especially important for 2D an 3D cases. 

The Eq.~(\ref{EQ:Cooperon/one}) describes
the WL conductance oscillations in the absence of magnetic flux 
 that occur in a ring at arbitrary SOI
 when $\beta$ is varied. 
When magnetic flux is varied, SOI leads to beatings of the 
AB oscillations.  

Consider now 2D systems, and assume the that the tensor $\beta$ has the form, 
appropriate for symmetric \cite{otherterms} GaAs/AlGaAs heterostructures: 
$\beta_{xx}=-\beta_{yy}=\beta_2$, 
where $z$ is the direction normal to the 2D plane 
(results for 2D Si with $\beta_{xy}=-\beta{yx}=\beta$ are the same). 
This type of term was first discussed by Altshuler et al.~\cite{Altshuler}.
Then, the solution for the Cooperon propagator reads: 
\begin{eqnarray}
C_{0,0} = 1/(1 - f), 
\label{EQ:coeff/a}\\
C_{1,0} = 1/(1 - f- 2g- 2h),
\label{EQ:coeff/b}\\
C_{1,\pm 1} = 1/(1-f-(3g+h)\pm\sqrt{t^2 + (g-h)^2 }),
\label{EQ:coeff/c}
\end{eqnarray}
Here the second index in Eqs.(5,6)
is the quantum number in the
representation diagonalizing the Cooperon,
\begin{eqnarray}
f = 1/\sqrt{1+ 2Dq^2\tau} 
\label{EQ:func/a}\\
g = (-1/\sqrt{1+ 2Dq^2\tau} + \sum_{\pm}1/\sqrt{a_{\pm}^2+ 2Dq^2\tau})/4 
\label{EQ:func/b}\\ 
h=\left[ -2/\sqrt{1+ 2Dq^2\tau} (\sqrt{1+ 2Dq^2\tau} + 1 +Dq^2\tau)\right.\nonumber \\  
+\left. \sum_{\pm}\!\frac{1}{\sqrt{a_{\pm}^2+ 2Dq^2\tau}(a_{\pm} + 
\sqrt{a_{\pm}^2+ 2Dq^2\tau})^2}\right]\!\frac{Dq^2\tau}{2}
\label{EQ:func/c} \\ 
t=\frac{ilq}{2} \sum_{\pm} \frac{(-1)^{(1 \pm 1)/2}}{
\sqrt{a_{\pm}^2+ 2Dq^2\tau}\left [ a_{\pm} + \sqrt{a_{\pm}^2+ 2Dq^2\tau} \right ]},
\label{EQ:func/d}
\end{eqnarray}
where $a_{\pm}=1 \pm 2i\beta m^{*}l = \pm 2i\Omega\tau$.

\begin{figure}[hbt]
\epsfxsize=\columnwidth
\epsfxsize=7.5truecm
   \centerline{\epsfbox{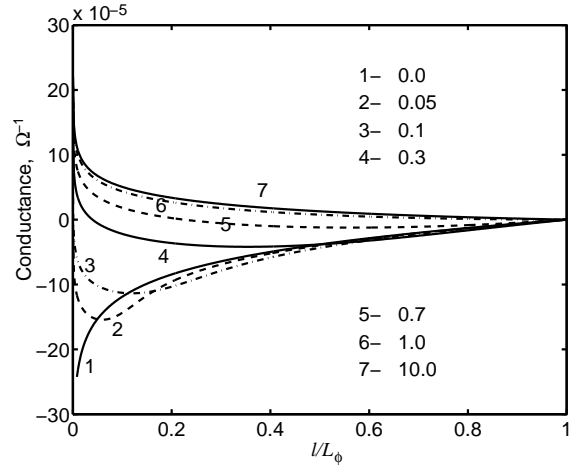}} 
\vskip+0.4truecm
\caption{The interference quantum correction 
to conductivity at various magnitudes of SOI strength $\Omega\tau$.}
\label{FIG:one}
\end{figure}
We now consider the consequences of Eqs.~(\ref{EQ:coeff/a}-\ref{EQ:func/d})
for interference corrections to the conductivity $G$, given by
\begin{equation}
-\frac{e^2 D\tau}{\pi}\int_{1/L_{\phi}}^{Q_0}\left(dQ\right)\left[- C_{00}
+C_{10}+C_{1,+1}+C_{1,-1}\right],
\label{EQ:COND}
\end{equation}
where $L_{\phi}$ is the phase-breaking-length, and $Q_0$ is the upper cutoff, 
usually~\cite{aim} regarded as being of order to $1/l$. The localization or antilocalization 
in weakly disordered conductors, the temperature ($T$), frequency and magnetic field ($H$) dependence 
of the conductivity is determined by $L_{\phi}$. On Fig.~\ref{FIG:one} I present the results
for the conductivity dependence on $L_{\phi}$ at various magnitudes of 
SOI strength $\Omega\tau$. In the absence of SOI (curve 1) one observes the weak localization. 
At small $\Omega\tau$ (curves 2-3) conductivity exhibits antilocalization, 
if SOI dephasing length 
$L_{so}=1/(\beta m^{*}) < L_{\phi}$, and weak localization in the opposite case. 
However, as $\Omega\tau$ approaches 1 (curves 4,5)\cite{remark}), 
the range of $L_{\phi}$ where electrons are localized 
diminishes. Finally, only antilocalization occurs at $\Omega\tau \ge 1$ (curves 6,7), 
because $l$ cannot exceed $L_{\phi}$.
Therefore, in contrast to random
SOI \cite{Deutcher}, 
as well as weak SOI in Eq.~(\ref{EQ:soi_term}), all studied 
earlier, single-particle corrections always lead to an 
{\it increase\/} in the conductivity at strong SOI. 
As $L_{\phi}^{-2}\propto T$ in 2D case~\cite{Alt_Ar}, 
Fig.~\ref{FIG:one} essentially represents 
($T^{1/2}$)-dependence of the conductivity. 
Similarly, for such 
$H$ perpendicular to 2D plane that magnetic length $L_H =(\hbar c/2eH)^{1/2} 
< L_{\phi}$, 
or such $\omega$ that $D/\omega < L_{\phi}^2$, 
Fig.~\ref{FIG:one} adequately describes the anomalous MR 
(conductance versus $H_{\perp}^{1/2}$) or 
the ($\omega^{1/2}$)-dependence of the interference correction to conductivity.

Antilocalization characterizing intereference corrections 
in 2D conductors in the whole range of temperatures, frequencies and orbital 
magnetic fields 
occurs due to entire strong-SOI-supression [described
by Eqs.(5-10)] of 
coherence of two electronic waves having total electron spin 1 and
moving along time-reversed paths. 
Moreover, such a suppression leads to the following behavior
of interference corrections to 
conductivity in magnetic field $H_{\parallel}$ lying in the 2D plane. 
 $H_{\parallel}$ influences both the singlet ($C_{00}$) and triplet ($C_{1j}$) sector of the 
 Cooperon propagator due to Zeemann effect. 
 For generic SOI all $C_{1j}$ components are mixed by $H_{\parallel}$.
At strong SOI $H_{\parallel}$ leads to increasing antilocalization at 
small magnetic fields, when it influences only the singlet component and is negligible 
for triplet states entirely suppressed by SOI. However,
at such magnetic fields that $g\nu H_{\parallel}\sim 1/\tau$ \cite{difference} 
$H_{\parallel}$ starts to suppress the triplet, and singlet and triplet 
contributions become comparable in magnitude. Thus, 
as the triplet contribution compensates the singlet one,
magnetic field dependence of the
conductivity weakens. 

\noindent {\it Interaction corrections to conductivity\/}:  
Quantum corrections to kinetic and thermodynamic quantities, due to 
electron-electron interactions in disordered conductors, have their origin in the 
enhancement of interactions between particles. 
The dominant contribution to this enhancement is due to electron diffusion 
leading to an increase of the interaction time and the effective 
interaction strength, for particles 
with small difference in momenta and energies, this process being described 
by corrections in the Diffuson channel \cite{Maki}.  
These corrections are 
not affected by the AB phase, but are influenced 
by the Zeeman interaction, magnetic impurities and SOI. 
As shown in Ref.~\cite{Lee_Alt}, positive MR arises because  
interaction of an electron and a hole with total 
spin 1 and spin-projection $\pm 1$ enhances the conductivity at $H=0$, 
but is suppressed due to the Zeeman interaction.  
The suppression of the electron-electron corrections to conductivity 
in the Diffuson channel by the weak SOI 
was discussed in Ref.~\cite{Alt_Ar}.  

We now discuss the implications of the SOI in Eq.~(\ref{EQ:soi_term}) 
for electron-electron interaction effects.
The effect of SOI on the Diffuson propagator is determined (at $H=0$)
by equations of the same form as Eqs. (4-6), with 
the net spin $S$ and spin projection $j$ describing the difference of electron spins 
(i.e. the total spin of electron and a hole). Strong SOI, therefore, entirely 
supresses contribution of the interaction of an electron and a hole with total 
spin 1 (refered below as the triplet Diffuson contribution) to the conductivity in 2D case. 
Under these conditions magnetic field has no effect on the 
interaction contribution to MR in Diffuson channel, as magnetic field does not
affect the contribution of the interaction of an electron and a hole with total 
spin 0 (refered as the singlet Diffuson contribution). Therefore, at strong SOI, MR
is determined by interference corrections. 
However, the temperature- and frequency-dependence of the conductivity are governed 
by the singlet Diffusin contribution 
 \cite{Alt_Ar}. The $T^{1/2}$-dependence of this singlet Diffuson correction  
 can be described well by the curve 1 on Fig.~\ref{FIG:one}, 
 but with the scale on the $y$-axis two times bigger and
$L_{\phi}$ on the $x$-axis substituted by $L_T \equiv\sqrt{D/T}$, in the range of $l/L_T\le 0.2$. 
(In this temperature range corrections from processes neglected at $T\tau \ll 1$ are not essential.)
  At strong SOI singlet Diffuson correction leads to the negative sign of 
the total quantum correction to conductivity which includes interaction and interference 
contributions. 
 
I now consider the oscillatory electron-electron interaction effects due to SOI. 
In quasi 1D case, SOI in (Eq.~\ref{EQ:soi_term}) leads to oscillations 
in ring-shaped samples of the interaction contributions to the 
conductance, and, in general,  all quantities affected 
by electron-electron interaction corrections in Diffuson channel. 
These oscillations with the variation of the 
SOI constant $\beta_1$ arise even in the absence of magnetic field due to  
SOI vector-potential ${\bf A}$. ${\bf A}$ affects the Diffuson propagator as given by 
the Eq.~(\ref{EQ:Cooperon/one}) and does not lead to 
SOI dephasing
in narrow \cite{narrow} 1D constrictions.
The triplet Diffuson contribution in which ${\bf A}$ manifests itself originates from 
Hartree contribution to the electron-electron quantum corrections. 
 I have calculated these corrections to the conductance of a ring.
The dominant contribution to the effect arises from terms
characterized by three diffusion poles in Hartree processes.
 If the temperature $T$ is sufficiently low then, for a ring of cross-sectional area 
$a^2$ and circumference $L \ll L_T$, the result of calculations 
of the oscillating contribution to conductance has the form
\begin{equation}
\delta\sigma^{{\rm osc}}
=\frac{e^2 L_T \lambda_1}{2^{3/2}\pi\hbar a^2 } 
 \sum_{n=1}^{\infty} e^{-\delta}
\left( sin{\delta}
-\cos{\delta}\right)\cos{n\eta}, 
\end{equation} 
where $\delta = nL/\sqrt{2}L_T$, $\eta =2\beta_1 m^* L$, and $\lambda_1$ 
(discussed in\cite{Alt_Ar,Finkelstein}) 
is the constant describing the interaction of an 
electron and a hole with total spin 1. Similar oscillations characterize the 
density of states and the thermodynamic potential. As the AB flux 
does not affect these electron-electron interaction contributions, and, at the same time, 
strong $H_{\perp}$ 
suppresses interference contributions, 
these oscillations may serve as an experimental
tool for investigating the triplet Diffuson corrections.  
The variation of $\beta_1$ leading to oscillations can be achieved 
by a gate voltage or uniaxial strain applied to a nanostructure. 

\noindent {\it Discussion of experimental settings\/}:
The SOI-effects considered in the present Letter can be observed 
in MR of 2D metallic samples at $E_F \tau \gg 1$. 
At strong SOI MR must be positive for all magnetic fields,
and the total quantum correction to the conductivity must be negative.
I now discuss the existing data of recent experiments 
\cite{Kravchenko}. One of the structures,
Si-$12b$, with high electron concentration $n_s =13.7\times 10^11 cm^{-2}$, 
$E_F =0.8meV$ ($10K$), and the conductivity 
$G=3.5e^2/(2\pi\hbar)$ at $T=2K$ is close to the range of parameters where 
the present consideration can be applied. 
This particular set of experimental data can be described in the following self-consistent picture.
The dimensionless conductivity  $G\sim E_F \tau + G_{\rm int} + G_{\rm ee}$,
where $G_{\rm int}$ is the interference contribution, and $G_{ee}$ is the interaction contribution.
$G_{ee}$ at such high temperatures ($T=2K$) is not logarithmic, 
as we estimate $T\tau \sim 0.8$ (because $E_F \tau \sim 3.2$ and $\tau=2.8\times 10^{-12}s$). 
Thus, $G_{ee}$ varies very slowly with $T$.
$G_{\rm int}$ is determined by intermediate SOI, 
as $\beta = 2.0\times 10^{-10} eV cm$ \cite{Pudalov1} and $\Omega\tau = 0.7$, and leads to an 
increase in the conductivity.
Assuming that $G_{\rm int}\sim G-E_F \tau\sim\ln{(L_{\phi}/l)}/\pi$ we obtain
 $l/L_{\phi}\sim 0.4$ which, in turn, gives  
$G_{\rm int}=0.36$ according to the curve 5 Fig.~\ref{FIG:one}. 
Considering the temperature dependence of interference correction given by
this curve we find $G\sim 5.5$ at $T=0.4K$, whereas in the experiment $G\sim 9$. 
As $\tau$ in this temperature range 
possibly increases, this values of $G$ 
seem to be in resonable agreement. Futhermore, at $T=0.3K$ the parameter $T\tau\sim 0.2$.
That is close to the region in which $G_{ee}$ becomes logarithmic 
and overcompensates $G_{\rm int}$. 
Thus, if this model is correct, a decrease in $T$ down to 0.1K must 
reveal a decrease in $G$ for this sample. 
Moreover, at $n_s$ above $13.7\times 10^{11} cm^{-2}$  the parameter $E_F \tau$ increases and 
they must reveal a decrease in $G$. Such studies at $n_s$ higher than $13.7\times 10^{11} cm^{-2}$
are more reliable because at $E_F \tau \sim 3$ the present theory is on the boundary of applicability. 
Such experiments, as well as a detailed study 
of MR at high $n_s$, have to confirm that at low T localization 
occurs in 2D metals. 

Although this Letter is not aimed at the analysis of those experiments in 
Refs.~\cite{Kravchenko,Simonian} 
in which $G\sim1$, I I would like to discuss the SOI strength in such a case. 
Its decrease estimated using the Drude model is not meaningful,
as neither Drude model nor the WL theory can be applied to 
this case. However, the renormalization of SOI strength with $G$ is possible 
and is important for a study of the regime $G\sim 1$ using 
scaling approach Ref.~\cite{Abrahams}.

\noindent {\it Concluding remarks\/}:
(i) The experimental tests proposed in this Letter for samples studied in 
Ref.~\cite{Kravchenko} may be helpful for elucidating 
the nature of the metallic state in Refs.~\cite{Kravchenko,MIT}. 
(ii) The experimental discovery of the AC 
oscillations in ring-shaped samples would bring the opportunity to distinguish 
the interference and interaction oscillatory contributions and to determine 
the electron-electron interaction constant $\lambda_1$.

\noindent {\it Acknowledgments\/}: 
I present my sincere gratitude to P.\ M.\ Goldbart for his attention to this work and 
numerous helpful discussions, and to I.\ L.\ Aleiner 
for helpful interesting discussions. Support by the U.S. Department of
Energy, Division of Materials Sciences under Award No. DEFG02-96ER45439 
is gratefully acknowledged.

 \end{multicols}

\begin{references}
\bibitem{strength} Although we allow for arbitrary $\Omega\tau$, we assume that
$\epsilon_{\rm F}\gg\hbar\Omega(p_{\rm F})$, so that SOI does not affect $\tau$.
\bibitem{narrow} It is assumed that narrow 1D constriction has transverse dimensions 
allowing for electron diffusion \cite{LGM}.
\bibitem{aim} Although Eqs.(4-6) are valid at arbitarary $ql$,
we do not aim to study the quasiballistic regime.
\bibitem{remark} At $\Omega\tau \sim 0.2\div 0.5$, the exact account of all orders in 
$ql$ appears to be especially important.
\bibitem{difference} In the case of relatively weak SOI \cite{Fukuyama} this condition reads 
$g\nu H_{\parallel}\sim 1/\tau_{so}$ and characteristic $H$ are much lower.
\bibitem{Pudalov1} This $\beta$ corresponds to the SOI gap $\Delta=3.6K$ 
in Ref.~\cite{Pudalov}. $\beta =1.8\times 10^{-9} eV cm$ in Ref.~\cite{Pudalov} is the overestimate.
\end{references}
 \end{document}